# Quantum splitter controlled by Rashba spin-orbit interaction.

I.A. Shelykh[1,2], N. G. Galkin[1], N.T. Bagraev[3]

[1]*LASMEA, CNRS, Université « Blaise-Pascal » Clermont Ferrand II, 24, av des Landais, 63177, Aubiere, France.*

[2]*St. Petersburg State Polytechnical University, 29, Politechnicheskaya, 195251, St-Petersburg, Russia.*

[3]*A.F.Ioffe Physicotechnical Institute, 194021, St. Petersburg, Russia*

We propose the mesoscopic device based on the Rashba spin orbit interaction (SOI) that contains a gated ballistic Aharonov-Bohm (AB) ring with incoming lead and two asymmetrically situated outgoing leads. The variations of the Rashba coupling parameter induced by the gate voltage applied to the AB ring is shown to cause the redistribution of the carrier flux between the outgoing leads and spin polarization of the outgoing currents, thus allowing the system to manifest the properties of the quantum splitter and spin filter.

**1. Introduction**

The role of the spin in the processes of the quantum transport has been in focus of both experimental and theoretical investigations in the last decade. The studies of the spin-orbit interaction (SOI) in semiconductor heterostructures have specifically attracted much interest to establish the principles of the spintronic devices that result from the precise manipulation and control of the spin of single electrons. Two mechanisms of the SOI appeared to be taken into account in mesoscopic systems: the so-called Dresselhaus SOI caused by the internal inversion asymmetry of the crystalline lattice and the Rashba SOI due to the structure inversion asymmetry. The latter mechanism was shown to be dominant in the Si-MOSFET [1] as well as in the InAs/GaSb, AlSb/InAs and GaAs/GaAlAs heterostructures [2-8]. The structure inversion asymmetry is important to be lifted by the external gate voltage, $V_g$, applied perpendicular to the structure interface thereby leading to the variations of the Rashba SOI coupling parameter $\alpha$. The possibility to tune the Rashba parameter $\alpha$ with $V_g$ was recently demonstrated experimentally for both electrons and holes in various materials [9-12] and was also the subject of a very detailed theoretical consideration [13-15].

The first spintronic device known as spin field effect transistor (FET) proposed in the pioneering work of Datta and Das [16] is based on the electron spin precession controlled by an external electric field via spin-orbit coupling. In frameworks of this proposal, the spin-polarized electrons are injected from a ferromagnetic source into the quasi-one-dimensional channel with a gate-controlled Rashba SOI, transferred through the channel in the ballistic regime, and finally detected at the ferromagnetic drain. The transmission amplitude of the electron depends on the relative alignment of its spin with the drain magnetization. The electron states inside the channel having the spin orientation collinear to the magnetization of the source are in general no longer eigenstates of the system because of the spin-orbit coupling in the region between the two ferromagnetic contacts. Thus, the spins of propagating electrons undergo precession in the effective magnetic field created by the Rashba spin-orbit coupling. Therefore, tuning of the Rashba parameter $\alpha$ with $V_g$ is able to result in the changes of the angle of the spin rotation in the region between the source and the drain thus leading to the variations of the current in the quasi-one-dimensional channel.

The main obstacle for the experimental realization of the Datta and Das spin transistor is the difficulty of the efficient spin injection from ferromagnetic contacts into semiconductors. The reported polarization degree of the spin injection is only about 1%, and the corresponding modulation of the conductance of the spin-FET is very small [17, 18]. In order to enhance the spin polarization degree, the atomically ordered interfaces between some ferromagnetic metals and semiconductors were recently proposed to act as ideal spin filters that transmit electrons of the only spin orientation [19, 20]. However, the experimental realization of near 100% spin injection remains still doubtful.

Recently, J. Nitta, F.E. Meijer and H. Takayanagi have proposed the spin-interference device which works without any ferromagnetic electrodes and external magnetic field [21, 22]. This device represents the ballistic Aharonov-Bohm (AB) ring covered by the gate electrode. The Rashba SOI in the AB ring induces the Aharonov- Casher phaseshift between the spin waves propagating in the clockwise and anticlockwise directions, which results in the large conductance modulation due to the interference of the spin wave functions [21, 23, 24].

Here we propose the spin interference device which does not need the spin-polarized carrier injection. This device represents a ballistic AB ring with one ingoing electrode, 0, and two asymmetrically situated outgoing electrodes, 1 and 2, shown in Fig. 1. The spin-dependent conductances $G_{01\pm}$ and $G_{02\pm}$ of this three-terminal device can be experimentally measured; where the index $\pm$ corresponds to the spin projection of the carrier on its wavevector in the outgoing leads. It should be noted that in the two-terminal devices with single propagating mode the outgoing current is unpolarized being due to the time- inversion symmetry [25]. However, using the multi-terminal structures where the electrons with opposite spins can be redistributed unequally between the outgoing leads, the effect of the spin filtering can be achieved [26,27].

In the ballistic regime the conductances $G_{01\pm}$ and $G_{02\pm}$ are determined by the phase relations between the waves propagating in the AB ring in clockwise and anticlockwise directions which are controlled by varying the Rashba parameter $\alpha$ as well as the value of the external magnetic field and the Fermi energy, $E_F$. The variations of $\alpha$ and the magnetic field value are shown to change efficiently the values of the total conductances $G_{01} = G_{01+} + G_{01-}$ and $G_{02} = G_{02+} + G_{02-}$, thus defining the redistribution of carriers between the outgoing leads. Besides, tuning of the Rashba parameter $\alpha$ appears to result in the variations of the polarization degree of the outgoing currents.

The present paper is organised as follows. In the second section we describe in detail the theoretical model of the quantum splitter based on the scattering matrix formalism and discuss qualitatively the spin filtering effect. In the third section the results of the numerical calculations of the conductances as a function of the external magnetic field and the Rashba coupling parameter are present. The influence of the intensity of the backscattering on the efficiency of the current redistribution between the outgoing leads and the spin polarization of the outgoing currents is studied. In conclusions, we summarize the main results of the work. In appendix, the analytical expressions for the amplitudes of the transmission into two outgoing leads are given.

## 2. Model

Here we consider the AB ring and the leads as being purely one-dimensional. This condition defines their crossection $L^2$ and the Fermi energy, $E_F = \frac{\pi^2}{m}\left(n_{2D} - \frac{\pi}{2L^2}\right)$, is small enough

$$\frac{mL^2 E_F}{\pi^2 \hbar^2} < 1 \qquad (1)$$

to provide propagation of all carriers by the only lowest one-dimensional quantum mode; where m is an effective mass of the carrier and $n_{2D}$ is the concentration of the 2D electron gas. It should be noted that the effects of the band mixing on the spin transport in the single quantum wire were analysed in detail in Refs 28, 29, but they are not relevant for the goal of the present work. Indeed, the single-channel devices are always preferable as compare to the multichannel devices because of much less effective spin relaxation [30]. Besides, the preparation of the single channel devices is now experimentally achievable [22, 31- 33].

In frameworks of the model presented, the case of the zero temperature is accented to provide the step-liked energy distribution of the carriers. Besides, the potentials of the two outgoing leads $V_{ds}$ are taken to be equal and small enough, $eV_{ds} \ll E_F$, so that the only carriers whose energy lies in the vicinity of the Fermi surface participate in the transport. The radius of the AB ring is taken to be much smaller than inelastic scattering length to satisfy the conditions of the ballistic transport. These conditions allow us to use the Landauer-Buttiker approach for the calculations of the conductances $G_{01\pm}, G_{02\pm}$ [34,35].

The conjunctions between the AB ring and leads are modelled by the quantum point contacts (QPCs) which provide the elastic scattering of the carriers. The QPCs are presumed to be identical and spin-independent. The latter assumption means that the spin of the carrier conserves during the passing through the QPCs. This model of the point contacts is used, because the experimental realisation of the spin-conserving QPCs has been shown to be possible [33], although other versions of the spin-dependent transport exist also [26]. Each QPC is characterised by the amplitude of the elastic backscattering of the carrier propagating inside the lead, $\sigma, |\sigma| < 1$, that is determined by the system geometry. The QPCs become completely transparent, if $\sigma = 0$.

It should be noted that the gate voltage $V_g$ applied to the AB ring effects not only on the Rashba parameter $\alpha$, but also on the wavevector of the carrier $k_F$ and the amplitude of the backscattering $\sigma$. However, we will as usually neglect below the changes introduced by the gate voltage $V_g$ in the values $k_F$ and $\sigma$, although this influence will be briefly discussed.

The uniform external magnetic field applied perpendicularly to the structure's interfacet is assumed to be weak enough to produce only the foregoing AB phase shift in the absence of any Zeeman splitting of the spin bands in both the leads and the AB ring. Besides, the spin projection is assumed to be conserved during the scattering of a particle by the QPC.

These assumptions allow to introduce the scattering matrix of the QPCs $S$ that is independent of the spin projections and able to relate the outgoing current amplitudes to the incoming current amplitudes on the QPCs (Fig. 1b) [36-38].

Since the only time-reversal invariant scattering at QPCs is taken into account, the amplitudes of the waves inside the AB ring $b_j, c_j, d_j$, $j=1,2$, and the amplitude of the reflected wave $a_2$ and transmitted waves $f_1, f_2$, which are shown in Fig. 1, are connected by the following set of nine linear equations:

$$\begin{pmatrix} b_1 \\ a_2 \\ d_2 \end{pmatrix} = \mathbf{S} \cdot \begin{pmatrix} b_2\tau_1 \\ 1 \\ d_1\tau_4 \end{pmatrix}, \quad \begin{pmatrix} c_2 \\ f_1 \\ d_1 \end{pmatrix} = \mathbf{S} \cdot \begin{pmatrix} c_1\tau_2 \\ 0 \\ d_2\tau_3 \end{pmatrix}, \quad \begin{pmatrix} c_1 \\ f_2 \\ b_2 \end{pmatrix} = \mathbf{S}' \cdot \begin{pmatrix} c_2\tau_1 \\ 0 \\ b_1\tau_2 \end{pmatrix}, \qquad (2)$$

where

$$\mathbf{S} = \begin{pmatrix} r & \varepsilon & t \\ \varepsilon & \sigma & \varepsilon \\ t & \varepsilon & r \end{pmatrix}. \quad (2a)$$

Here $r$ and $t$ are reflection and transmission amplitudes of the QPCs inside the AB ring. $\sigma$ is the reflection amplitude from the lead to itself, whereas $\varepsilon$ is the transmission amplitude from a lead to the AB ring or from the AB ring to a lead. Simplifying, the scattering amplitudes $r, t, \sigma$ and $\varepsilon$ are assumed to be real numbers, the addition of the imaginary parts does not qualitatively change the results. These parameters depend on the properties of the conjunction, in particular on the band mismatch between the leads and the AB ring that can be electrically induced by the gate voltage $V_g$. $\tau_j$ describe the phaseshifts of the waves propagating between the different QPCs. The corresponding expressions are given below (see formulae 10- 10c)

The number of independent matrix elements can be reduced, as the scattering matrix is unitary owing to the conservation of the flux. Following Buttiker et al. [38], the parameterisation of the scattering matrix reads:

$$r = \frac{\lambda_1 + \sigma}{2} \quad (3)$$

$$t = \frac{-\lambda_1 + \sigma}{2} \quad (3a)$$

$$\varepsilon = \lambda_2 \sqrt{(1-\sigma^2)/2} \quad (3b)$$

where $\lambda_{1,2}$ are either 1 or -1. Therefore the effect of the QPCs on the scattering of a particle in the AB ring appears to be defined only by the parameter $\sigma$.

In order to analyse the spin-dependent transport of carriers inside the gated AB ring, the SOI Hamiltonian for the one-dimensional channel will be taken into account [25, 39, 40]

$$H = \frac{\hbar^2}{2ma^2}\left[i\frac{d}{d\varphi} + m\alpha a\left(\sigma_y \cos\varphi + \sigma_x \sin\varphi\right)\right]^2 \quad (4)$$

where $\varphi$ is a rotation angle, $\beta = 2m\alpha a$ is a dimensionless of the SOI constant in the AB ring. The eigenstates of the SOI Hamiltonian read

$$\psi_1 = \frac{e^{ik_+ a\varphi}}{\sqrt{1+\xi^2}}\begin{pmatrix} i\xi \\ e^{-i\varphi} \end{pmatrix} \quad (5)$$

$$\psi_2 = \frac{e^{ik_- a\varphi}}{\sqrt{1+\xi^2}}\begin{pmatrix} 1 \\ i\xi e^{-i\varphi} \end{pmatrix} \quad (6)$$

where $\xi = 2m\alpha a/\left(1+\sqrt{1+(2m\alpha a)^2}\right)$. The sign of the wavenumbers $k_\pm$ determines the direction of the motion, whereas their absolute values can be found from the following equation [25]:

$$\left(\frac{2mE_F}{\hbar^2}-k_{\pm}^2\right)\left[\frac{2mE_F}{\hbar^2}-\left(k_{\pm}-1/a\right)^2\right]-4m^2\alpha^2\left(k_{\pm}-1/2a\right)^2=0 \quad (7)$$

The eigenstates $\psi_1$ and $\psi_2$ correspond to the two orthogonal spin orientations. If the AB ring radius is large, so that $k_{\pm} \gg 1/a$, the spin of the two eigenstates is oriented in plane of the AB ring, towards or from the center. This result is easily understandable from the classical point of view. Indeed, neglecting the AB ring curvature, the effective magnetic field created by the Rashba SOI is given by the vector product of the external electric field and the carrier wavevector,

$$\mathbf{B}_{eff} = \frac{\alpha}{g_B \mu_B}[\mathbf{k} \times \mathbf{e}_z] \quad (8)$$

Thus, in frameworks of the model presented, the only the radial component of the effective magnetic field exists, which can be positive or negative being dependent on the direction of the carrier motion and the sign of the Rashba parameter $\alpha$ controlled by the sign of $V_g$. In the adiabatic approximation, which will be used in the further calculations, the spin of the carrier moving inside the AB ring is either parallel or antiparallel to the effective magnetic field, so that the carriers with the same radial component of the spin propagating in opposite directions have the different wavenumbers. Indeed, after entering to the AB ring through the ingoing lead, the carrier with a spin collinear with a wavevector can propagate in the clockwise or anticlockwise direction, correspondingly with parallel and antiparallel spin to the effective magnetic field $\mathbf{B}_{eff}$ (or vice versa, depending on the sign of $V_g$). Since the carrier energy should be the same in both cases, $E_F + eV_g = \frac{\hbar^2 k_{\pm}^2}{2m} \mp g_B \mu_B B_{eff}$, the wave numbers $k_+$ and $k_-$ of the electron propagating in the clockwise and anticlockwise directions are given by the following formula

$$k_{\pm} = \pm\frac{m\alpha}{\hbar^2}+\sqrt{\left(\frac{m}{\hbar^2}\right)\left(\frac{m\alpha^2}{\hbar^2}+2(E_F+eV_g)\right)} \quad (9)$$

Where $eV_g$ is an electrically induced band mismatch between the leads and the AB ring.

Thus, for the electron with spin oriented towards the center of the AB ring the phaseshifts $\tau_i$ can be calculated as:

$$\tau_1 = \exp\left[i\frac{\pi}{2}\left(k_+ a - \Phi/\Phi_0 + \frac{1}{2}\right)\right] \quad (10)$$

$$\tau_2 = \exp\left[i\frac{\pi}{2}\left(k_- a + \Phi/\Phi_0 - \frac{1}{2}\right)\right] \quad (10a)$$

$$\tau_3 = \exp\left[i\pi\left(k_+ a - \Phi/\Phi_0 + \frac{1}{2}\right)\right] \quad (10b)$$

$$\tau_4 = \exp\left[i\pi\left(k_- a + \Phi/\Phi_0 - \frac{1}{2}\right)\right] \qquad (10c)$$

Where $\Phi = \pi a^2 B$ is the magnetic flux through the AB ring, as well as $a$ is the radius of the AB ring and $\Phi_0$ is an elementary flux quantum. The first term at the argument of the exponent corresponds to the Aharonov-Casher phase, the second to the Aharonov-Bohm phase, the third to the geometrical Berry phase [21, 41-43]. Finally, mutual orientation of the spin and the effective magnetic field is opposite to the discussed above, when the spin of the carrier in the ingoing lead is opposite to its wave vector. In this case the values of $k_+$ and $k_-$ should be interchanged in the calculations of the phaseshifts (see formulae (10)-(10c)).

The phase shifts are seen to be dependent on the Rashba parameter $\alpha$ and consequently the modulation of the conductances is observed not only in the AB oscillations as a function of the magnetic field, but also in the Aharonov-Casher (AC) oscillations [44] as a function of the gate voltage. Besides, the phaseshifts are different for the carriers with opposite spin projections thereby giving rise to the different values of the conductances $G_{01,2+}$ and $G_{01,2-}$, so that the device considered may demonstrate the properties of the spin filter. The spin filters based on the AB rings were already proposed in the literature [45, 46], but their structure changed sufficiently from considered in this paper. Quite importantly, both of them need external magnetic field, while in the structure we consider the filtering effect that is induced only by electrical field.

The equations (2)-(3) lead to the analytical expressions for the amplitudes of the transmission into the two outgoing leads which are given in the appendix. These expressions permit to calculate the conductances of the AB ring:

$$G_{01,2\pm} = 2\frac{e^2}{h}|f_{01,2\pm}|^2 \qquad (11)$$

which appear to be dependent on the energy of a particle, the value of the external magnetic field and the Rashba parameter $\alpha$. The formula (11) allows the calculation of the total conductances of the two outgoing leads and corresponding polarization degrees of the outgoing currents:

$$G_{01,2} = G_{01,2+} + G_{01,2-} \qquad (12)$$

$$P_z^{1,2} = \frac{|f_{01,2+}|^2 - |f_{01,2-}|^2}{|f_{01,2+}|^2 + |f_{01,2-}|^2} \qquad (12a)$$

$$P_x^{1,2} = \frac{\operatorname{Re} f_{01,2+} f_{01,2-}^*}{|f_{01,2+}|^2 + |f_{01,2-}|^2} \qquad (12b)$$

$$P_y^{1,2} = \frac{\operatorname{Im} f_{01,2+} f_{01,2-}^*}{|f_{01,2+}|^2 + |f_{01,2-}|^2} \qquad (12c)$$

Where z direction coincides with the direction of the corresponding outgoing lead, and $P_x^{1,2} = P_y^{1,2} \equiv 0$, because in the adiabatic approximation the spin of the carrier in the AB ring is either parallel or antiparallel to the effective magnetic field. But the value of $P_z^{1,2}$ is nonzero and is determined by the value of the Rashba parameter $\alpha$.

## 3. Numerical results

The results of numerical calculations of the conductances depending on the external fields are shown in Figs. 2-5. We have considered the model of the GaAs related heterostructure with the following values of the parameters: $m = 0.06 \cdot m_e$, $E_F = 0.5\,\text{meV}$, $a = 1.0\,\mu\text{m}$.

The figure 2 shows the dependences of the conductances $G_{01}$ and $G_{02}$ on the external magnetic field. We have neglected the Zeeman spin splitting inside the AB ring and the leads, so that the conductances $G_{01}$ and $G_{02}$ are spin-independent. If $\sigma = 0$ and the scattering on the QPCs is absent (see Fig.2-solid line), they have a clear periodical pattern with a period equal to the quantum of the flux (h/e AB oscillations). The oscillations of the $G_{01}$ and $G_{02}$ conductances are phaseshifted, and thus the magnetic field can be used for redistribution of the ballistic current between the two outgoing leads, leading to the nonzero value of the parameter

$$Q = \frac{G_{01} - G_{02}}{G_{01} + G_{02}} \tag{13}$$

The form of the oscillations changes dramatically, if the backscattering of the carriers on the conjunctions between the AB ring and the leads is taken into account. Although the period of the oscillations rests the same, each of the peaks in the conductances is now split into two asymmetric peaks (see Fig.2, dash line). Consequently, the harmonics with the frequency corresponding to the half of the flux quantum becomes dominant (h/2e Aaronov-Altshuler-Spivak oscillations) [47]. This change of the period of the oscillation seems to be due to the increase of the effective path length for the carriers inside the AB ring which is induced by the enhancement of the amplitudes of the backscattering on the QPCs and reveal the importance of the correct account of the carrier's round trips.

The dependence of the conductances on the Rashba parameter $\alpha$ is shown in Fig.3. The value of the external magnetic field was taken zero, and the backscattering on the QPCs was neglected. The total conductances of the both outgoing leads $G_{01}$ and $G_{02}$ are seen to reveal the AC-type oscillations. As the AB oscillations discussed earlier, the AC oscillations are phaseshifted, and thus tuning of the Rashba parameter $\alpha$ allows the control of the direction of the carrier's motion after passing the gated AB ring (Fig. 3d). Besides, the conductances $G_{01}$ and $G_{02}$ become to be spin-dependent, and the outgoing currents are spin-polarized. The polarization degree as a function of the Rashba parameter $\alpha$ is shown at the Fig. 3c.

The amplitude and the form of the AC oscillations of the conductance strongly depend on the backscattering amplitude $\sigma_{1,2}$ (see Fig. 4). Each conductance peak doubles due to the increase of the probability of the round trips in the system. The increase of the backscattering amplitude is also seen to decrease the difference of the conductances of two spin components, thus suppressing the spin filtering effect.

The calculation presented above was carried out in the assumption that the Rashba parameter $\alpha$ can be tuned independently of $\sigma$ and the band bottom inside the AB ring. However, all these parameters were noted to be dependent simultaneously on the gate voltage $V_g$ that introduces the band mismatch between the AB ring and the leads thereby increasing the probabilities of the backscattering at the QPCs. In the approximation $eV_g < E_F$ the effect of the gate voltage $V_g$ on the value of $\sigma$ can be roughly estimated as (see Appendix 2)

$$\sigma \approx \frac{eV_g}{4E_F} \qquad (14)$$

Besides, according to the equations (7, 9) the gate voltage $V_g$ affects directly the values of $k_+, k_-$. Analysing the Datta and Das device, this effect was shown to give rise to the appearance of the additional resonances in the dependence of the conductance on the gate voltage $V_g$ [48, 49]. The width of these resonances is however much smaller than the width of the AC oscillations provided by the modulation of the Rashba parameter $\alpha$. Therefore these additional resonances seem to be observed only at the extremely low temperatures (less than 1K).

The corresponding dependence of the conductance on the gate voltage is shown in Fig.5. The value of the external magnetic field was taken zero. The AC oscillations of the $G_{01}$ and $G_{02}$ conductances in the three-terminal device studied are seen to be strongly asymmetric. The pronounced asymmetric resonances whose shape resembles Fano resonances is of interest to be found at the voltages about 2,33 and 2,48 mV. Beginnings of these resonances in three-terminal devices are rather surprised, because so far the Fano resonance structure caused by the interference of the continuous and bound states is revealed by the quantum dot strongly coupled with a quantum wire [50]. This effect is required to be studied in details, specifically on the value of the backscatterring amplitude, $\sigma_{1,2}$.

## 4. Conclusions

In conclusion, we have shown that the gated ballistic AB ring with three asymmetrically situated electrodes is the spintronic device that is able to demonstrate the properties of both the quantum splitter and spin filter. The conductances and spin polarization degrees of the outgoing currents in this three-terminal device are predicted to be depend on the external magnetic field and the gate voltage. Consequently, tuning of one of these parameters allows the efficient control of the redistribution of the current between the two outgoing leads and its spin polarization.

## 5. Appendix 1

The amplitudes of the transmission into the two outgoing leads are determined from equations (2)-(3). After simple, but tiresome algebraic calculations we can be obtain

$$f_1 = \frac{A}{C}, \quad f_2 = -\frac{B}{C},$$

where

$$A = \varepsilon_1^2 \left[ \begin{array}{l} (r-t)^2 t'^2 \tau_1^2 \tau_2^2 \tau_3 - (1 - rr'\tau_1\tau_2 + t't\tau_1\tau_2)^2 \tau_3 + \\ + t'\tau_2^2 \left( -1 + r^2 \tau_3\tau_4 - 2tr\tau_3\tau_4 + t^2 \tau_3\tau_4 \right) \end{array} \right]$$

$$B = \varepsilon_1 \varepsilon_2 \left\{ \begin{array}{l} t\tau_1\tau_3 + \tau_2 \left[ 1 + r'r^2 \tau_1^2 \tau_3 - t^2 t' \tau_1^2 \tau_3 - r^2 \tau_3\tau_4 + rt\tau_3 \left( -r'\tau_1^2 + t'\tau_1^2 + \tau_4 \right) \right] + \\ + (r'-t')\tau_1\tau_2^2 \left[ r^3 \tau_3\tau_4 - r^2 t\tau_3\tau_4 + t^3 \tau_3\tau_4 - r(1 + t^2 \tau_3\tau_4) \right] \end{array} \right\}$$

$$C = (r^4 + t^4)(r'^2 - t'^2)\tau_1^2 \tau_2^2 \tau_3 \tau_4 - 2r^3 r' \tau_1 \tau_2 \tau_3 \tau_4 + t^2 t'(\tau_1^2 \tau_3 + \tau_2^2 \tau_4) + \\ + 2rr'\tau_1\tau_2(1 + t^2\tau_3\tau_4) + t^2 \left[ \tau_3\tau_4 + (t'^2 - r'^2)\tau_1^2 \tau_2^2 (1 + 2t^2 \tau_3\tau_4) \right] - 1$$

## 6. Appendix 2

The formula (14) for the amplitude of the backscattering due to the band mismatch between the ring and the lead can be obtained as follows. Imagine that the gate is applied to the ring and short neighboring part of the lead. The carrier then undergoes backscattering on the potential mismatch within the lead, the amplitude of which can be simply calculated as

$$\sigma = \frac{\sqrt{1 + eV_g/E_F} - 1}{\sqrt{1 + eV_g/E_F} + 1}$$

In the limit of the weak gate, when $eV_g/E_F \ll 1$ the Taylor decomposition of this formula gives (14)

Figure captions.

**Fig. 1.** Schematic view of a spin-interference device [15, 24] that is based on the AB ring connected with two one-dimensional leads by QPCs and covered by the gate electrode that controls the Rashba SOI with the amplitudes of travelling electronic waves.

**Fig. 2**. The dependences of the conductances $G_{01}$ and $G_{02}$ and redistribution parameter Q on the external magnetic field $\Phi/\Phi_0$. The conductances as a function of the magnetic field show the pronounced Aharonov- Bohm (solid lines) and Aharonov- Altshuler- Spivak oscillations (dashed lines). The asymmetrical placing of the two outgoing leads allows the redistribution of the outgoing current, charachterised by the parameter Q.

**Fig. 3.** Conductances $G_{01,2+}, G_{01,2-}$, $G_{01,02}$, the spin polarization degree and the redistribution parameter Q as a function of the Rashba coupling parameter $\alpha$. The backscattering on the QPCs is absent, $\sigma_{1,2} = 0$. The conductances show the Aharonov- Casher oscillations. The asymmetry of the structure allows the redistribution of the outgoing current and its spin polarization by tuning of the Rashba parameter.

**Fig. 4.** Conductance $G_{02\pm}, G_{02}$ as a function of the Rashba coupling parameter $\alpha$. The amplitude of the backscattering on the contacts is $\sigma_{1,2} = 0.8$. The resonances of the conductance are splitted due to the account of the round trips of the carrier inside the ring. The inset shows the spin polarization degree of the outgoing currents, which is suppserred comparing with tha case when the backscattering is absent.

**Fig. 5**. The dependences of the $G_{01}$ and $G_{02}$ conductances on the gate voltage. The scatterring on the QPCs $\sigma_{1,2}$, band bottom inside the AB ring and the Rashba coupling parameter $\alpha$ are dependent on the gate voltage $V_g$. The pronounced Fano resonances are seen at the energies 0.02, 008 and 0.24 mV. The inset shows the spin polarization degree of the outgoing currents.

Fig.1

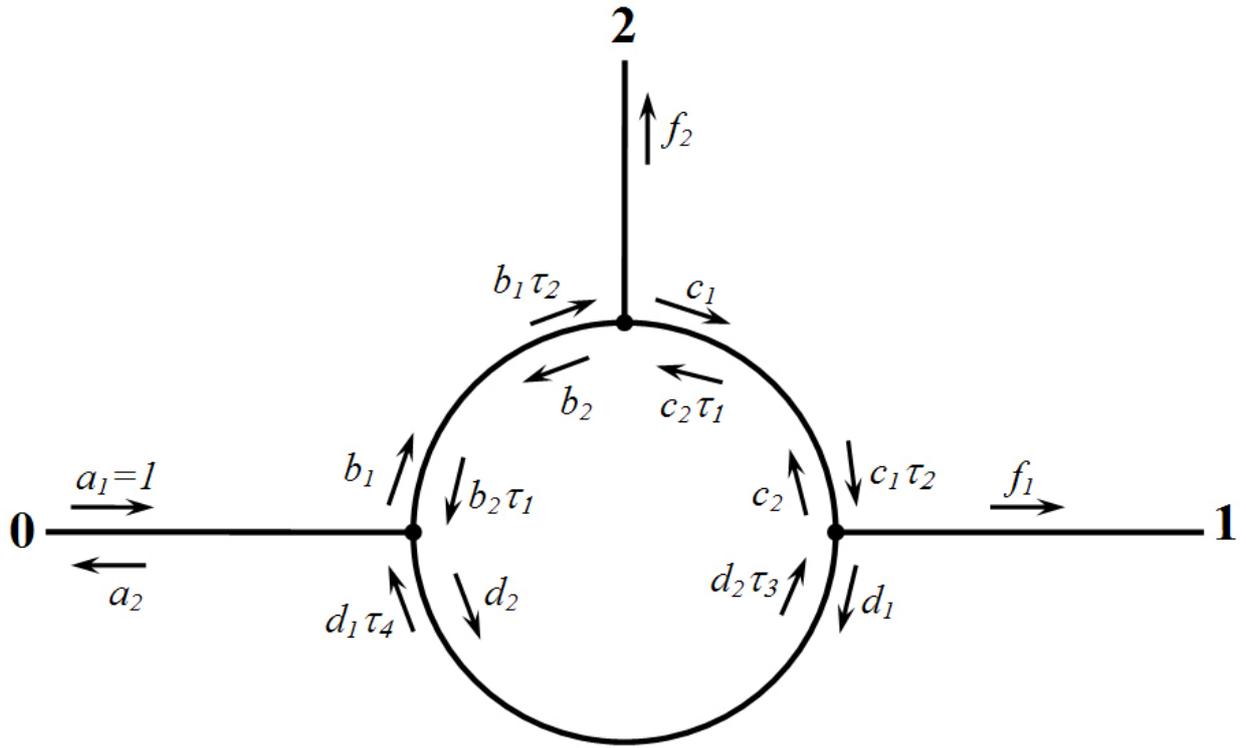

Fig.2

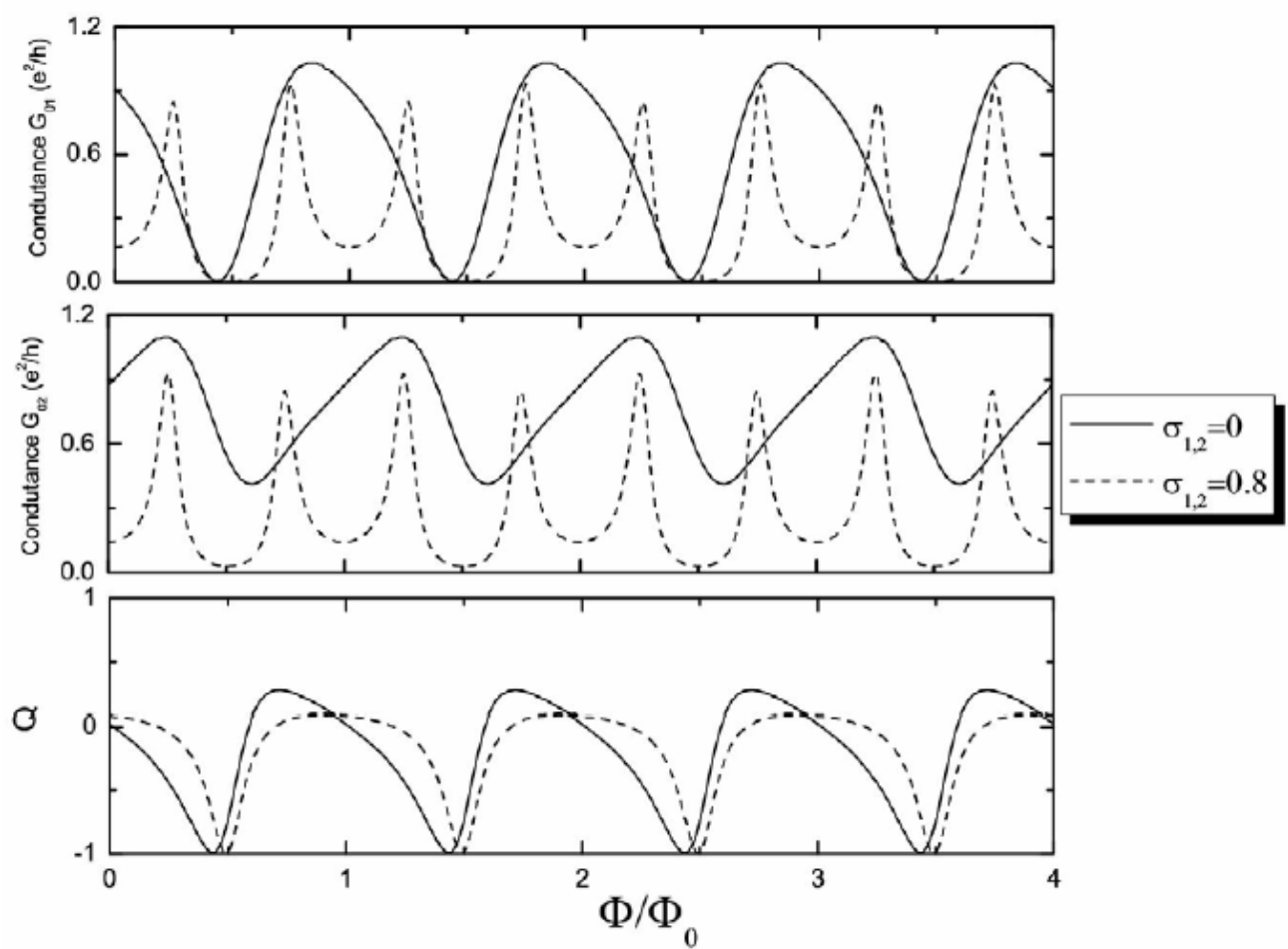

Fig.3

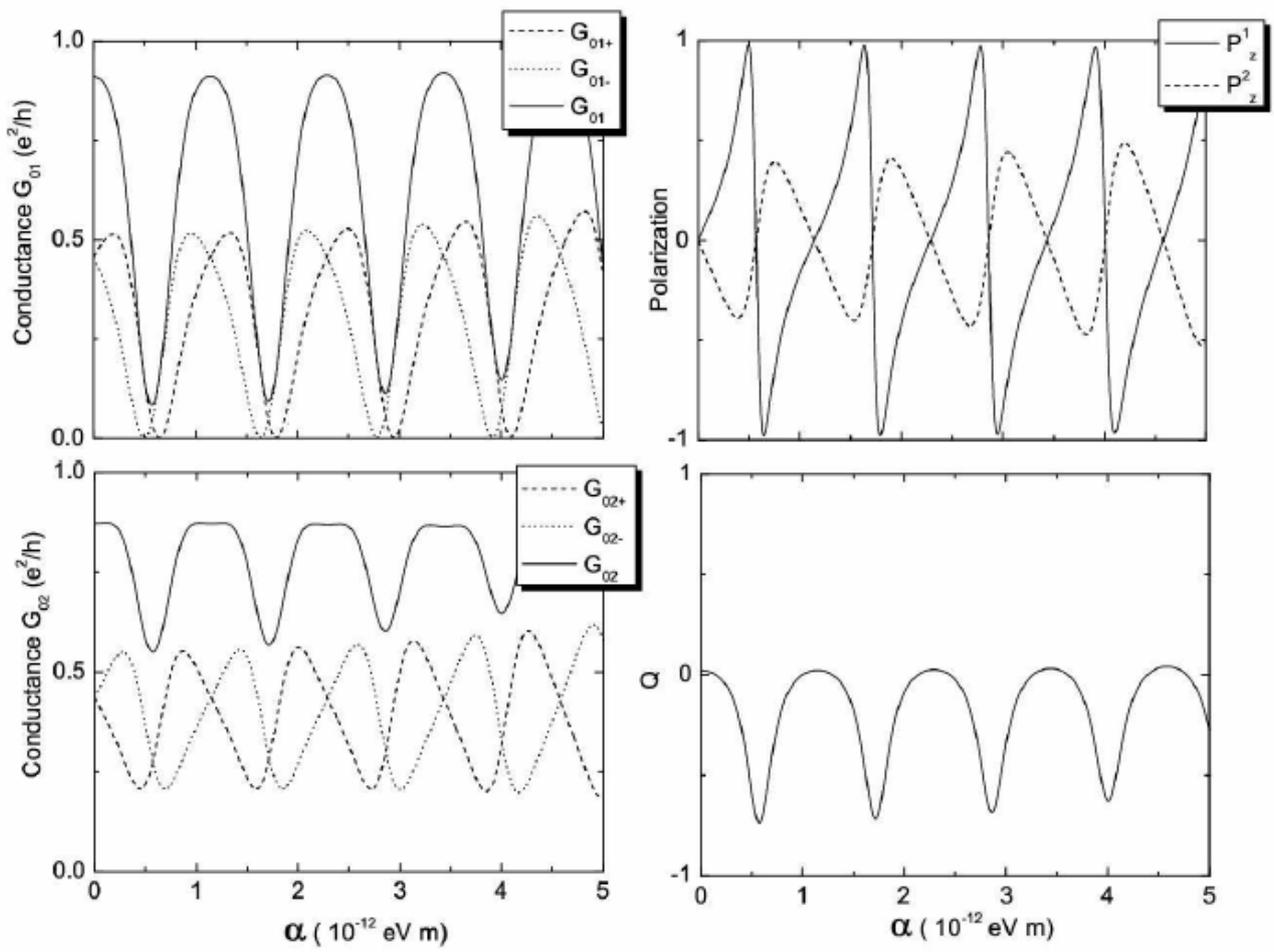

Fig.4

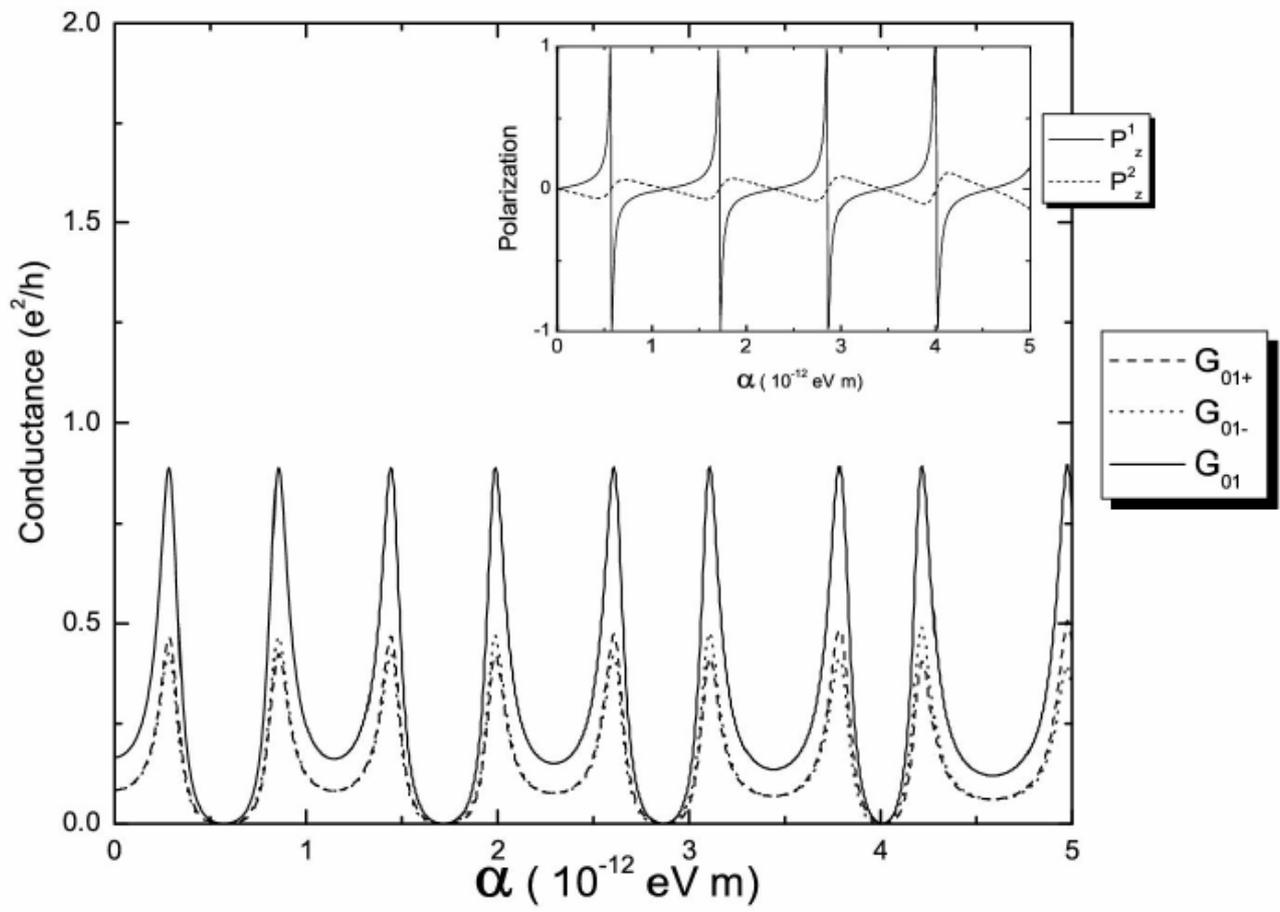

Fig.5

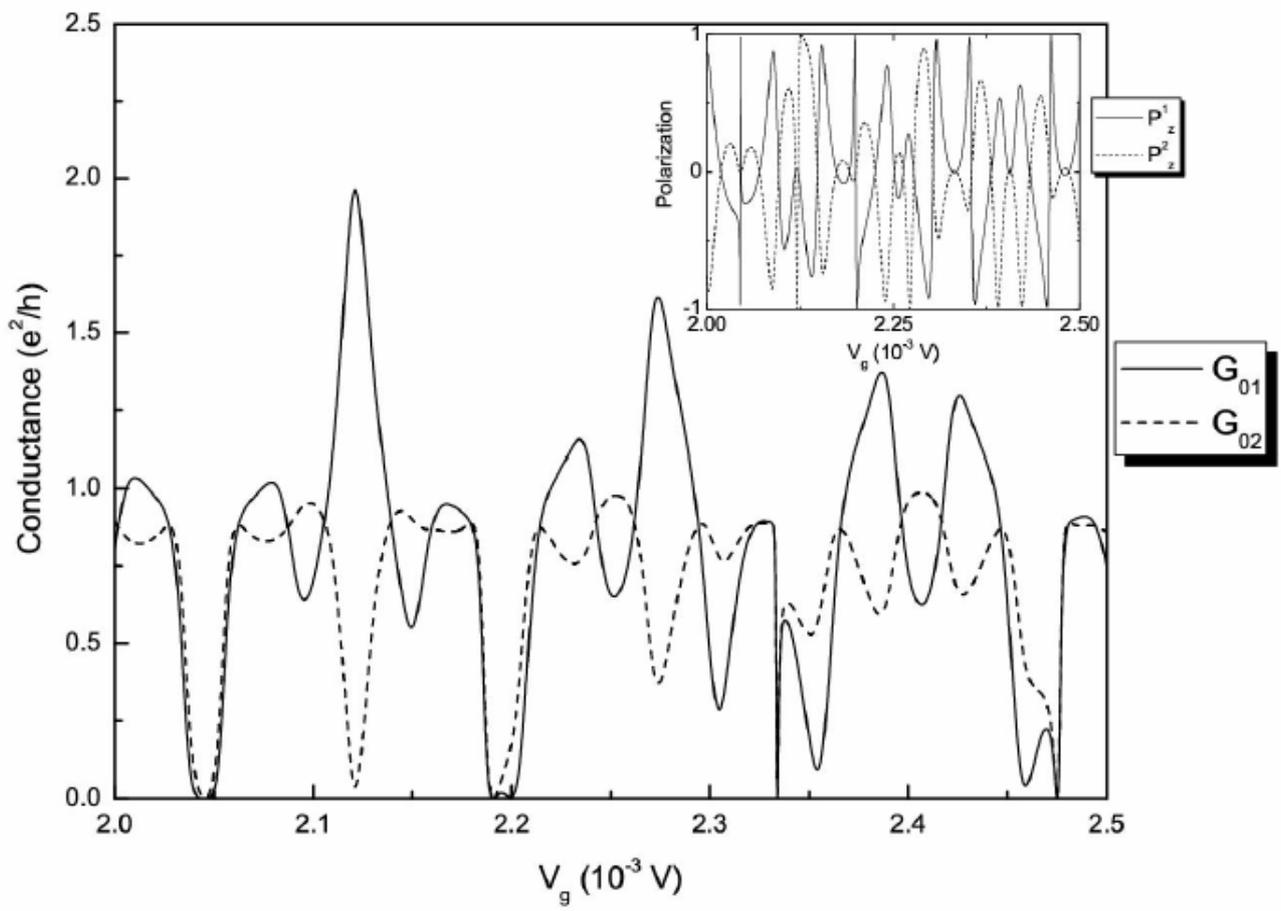